\documentclass{article}
\usepackage{amsmath,amsfonts,amssymb,amsthm,graphicx}
\pdfoutput=1
\usepackage[preprint]{spconf}
\usepackage{url}
\usepackage{hyperref}
\usepackage{caption}
\usepackage{balance}
\usepackage{bm}

\def\CC{{\mathbb C}}
\def\RR{{\mathbb R}}
\usepackage{cite}
\usepackage{xcolor}
\usepackage{multirow}
\usepackage{etoolbox,siunitx}
\usepackage{svg}
\robustify\bfseries
\usepackage{booktabs}

\usepackage{enumitem}

\usepackage{mathtools}
\usepackage{siunitx} 
\usepackage{tabularx, booktabs} 

\let\OLDthebibliography\thebibliography
\renewcommand\thebibliography[1]{
  \OLDthebibliography{#1}
  \setlength{\parskip}{.5pt}
  \setlength{\itemsep}{.5pt plus 0.3ex}
}

\newcommand\blfootnote[1]{%
  \begingroup
  \renewcommand\thefootnote{}\footnote{#1}%
  \addtocounter{footnote}{-1}%
  \endgroup
}

\makeatletter
\def\blfootnote{\gdef\@thefnmark{}\@footnotetext}
\makeatother

\title{GLA-Grad: A Griffin-Lim Extended Waveform Generation Diffusion Model}
 
\name{Haocheng Liu$^{1}$,
    \quad Teysir Baoueb$^{1}$, 
    \quad Mathieu Fontaine$^{1}$, 
    \quad Jonathan Le Roux$^{2}$, 
    \quad Gaël Richard$^{1}$}
  
\address{$^{1}$LTCI, T\'el\'ecom Paris, Institut polytechnique de Paris, France\\
      $^{2}$Mitsubishi Electric Research Laboratories (MERL), Cambridge, MA, USA}

\usepackage{amsmath,amsfonts,amssymb,bm}

\newcommand{\Tr}{\mathsf{T}}

\newcommand{\mbC}{\mathbf{C}}
\newcommand{\mbD}{\mathbf{D}}

\newcommand{\mbI}{\mathbf{I}}

\newcommand{\mbL}{\mathbf{L}}
\newcommand{\mbM}{\mathbf{M}}

\newcommand{\mbS}{\mathbf{S}}
\newcommand{\mbT}{\mathbf{T}}

\newcommand{\mbX}{\mathbf{X}}

\newcommand{\mby}{\mathbf{y}}
\newcommand{\mbz}{\mathbf{z}}

\newcommand{\mbep}{\boldsymbol{\epsilon}}

\newcommand{\mbSi}{\boldsymbol{\Sigma}}

\newcommand{\mbPh}{\boldsymbol{\Phi}}

\copyrightnotice{\copyright 2024 IEEE. Personal use of this material is permitted. Permission from IEEE must be obtained for all other uses, in any current or future media, including reprinting/republishing this material for advertising or promotional purposes, creating new collective works, for resale or redistribution to servers or lists, or reuse of any copyrighted component of this work in other works.}
\begin{document}

\ninept
\maketitle

\begin{abstract}
Diffusion models are receiving a growing interest for a variety of signal generation tasks such as speech or music synthesis. WaveGrad, for example, is a successful diffusion model that conditionally uses the mel spectrogram to guide a diffusion process for the generation of high-fidelity audio.  However, such models face important challenges concerning the noise diffusion process for training and inference, and they have difficulty generating high-quality speech for speakers that were not seen during training.  
With the aim of minimizing the conditioning error and increasing the efficiency of the noise diffusion process, we propose in this paper a new scheme called GLA-Grad, which consists in introducing a phase recovery algorithm such as the Griffin-Lim algorithm (GLA) at each step of the regular diffusion process.  Furthermore, it can be directly applied to an already-trained waveform generation model, without additional training or fine-tuning. We show that our algorithm outperforms state-of-the-art diffusion models for speech generation, especially when generating speech for a previously unseen target speaker.
\end{abstract}

\section{Introduction}
\blfootnote{This work was funded by the ANR Project SAROUMANE (ANR-22-CE23-0011) and by the European Union (ERC, HI-Audio, 101052978). Views and opinions expressed are however those of the author(s) only and do not necessarily reflect those of the European Union or the European Research Council. Neither the European Union nor the granting authority can be held responsible for them.}

In the era of deep learning, generating high-fidelity audio data is a major research topic for improving the performance of machine listening algorithms.
Apart from audio synthesis for listening applications, it can also be seen as an important asset for data augmentation in unbalanced settings, such as in sound event detection, speech processing, or music information retrieval-related tasks.

Speech generation has witnessed the utilization of various architectures and input methods to produce high-quality audio speech.
Notable algorithms in speech generation encompass text-to-speech (TTS) variational autoencoder \cite{lu21d_interspeech}, as well as convolutional neural network (CNN) architectures operating directly on raw audio 
such as Wavenet \cite{vanwavenet}. 
Mel spectrograms serve as an alternative input for speech generation, with such examples as Waveglow \cite{prenger2019waveglow}, based on normalizing flow \cite{rezende2015variational}, and HiFi-GAN \cite{kong2020hifi}, based on a generative adversarial network (GAN) \cite{goodfellow2014generative}.

Diffusion models \cite{ho2020denoising, song2021scorebased} have recently emerged as highly advanced deep generative models where a stochastic flow to solve differential equations is governed by a fixed noise schedule.
They have surpassed the long-standing dominance of  GANs in the field of image synthesis \cite{huang2018introduction, liu2020generative, dhariwal2021diffusion}.
Diffusion models have also been considered for various speech-related tasks such as speech enhancement \cite{lemercier2023storm, richter2023speech, lu2022conditional} or speech separation \cite{scheibler2023diffusion, chen2023sepdiff}.
Diffusion-based speech generation methods that use conditioning on mel spectrograms have also been proposed, notably Wavegrad \cite{chen2020wavegrad} and DiffWave \cite{kong2020diffwave}. Building upon DiffWave,  PriorGrad \cite{lee2021priorgrad} replaces the standard Gaussian prior noise assumption by an adaptive prior, while SpecGrad \cite{koizumi2022specgrad} applies constraints to ensure the time-varying spectral envelope aligns with the log-mel spectrogram. 

Nonetheless, diffusion models face an ongoing challenge in efficient training. Their stability relies heavily on large datasets to ensure the stability of the diffusion process through the iterations \cite{wang2023patch}.
An essential component is the noise schedule governing the reverse process for generating speech. Recent advances such as FastDiff \cite{huang2022fastdiff} propose to predict the noise schedule dynamically throughout iterations to reduce their number and enhance the signal generation.
Furthermore, the versatility of speech signals presents domain adaptation limitations with a diffusion model framework \cite{farahani2020brief}. These limitations result in poor generalization 
to unseen speakers in speech generation scenarios.

\begin{figure}[t]
\centering
\includegraphics[width=0.98\columnwidth]{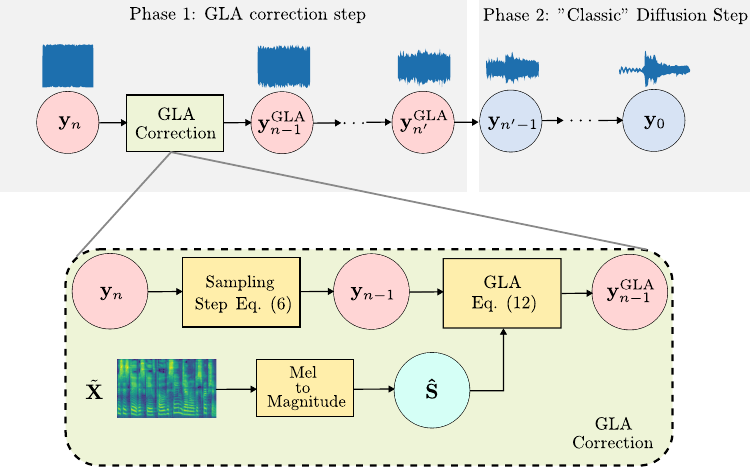}
\caption{Griffin-Lim corrected diffusion model}
\vspace{-.3cm}
\label{fig:method}
\end{figure}

Our paper addresses the latest drawback which, to the best of our knowledge, has not been explored within the diffusion model framework for speech generation. In this article, we introduce a straightforward extension to WaveGrad, aiming to rectify conditioning errors on the mel spectrogram during the reverse process iterations. 
As illustrated in Fig.~\ref{fig:method}, we use the Griffin-Lim algorithm (GLA)  \cite{griffin1984signal} to partially reconcile the phase of the short-time Fourier transform of the current sample in the diffusion process with the magnitude spectrogram obtained as pseudo-inverse of the conditioning mel spectrogram.
Our objective is to ensure consistency in the reverse  process, even when the user-provided signal significantly differs from those in the training set.

The achieved scores demonstrate GLA-Grad's competitiveness with WaveGrad and SpecGrad when using a single-speaker dataset.
However, GLA-Grad significantly outperforms them as the dataset's speaker count increases, highlighting its superior generalization capability across multiple speakers.
Additionally, experiments involving unseen speakers during training demonstrate GLA-Grad's effectiveness in addressing domain adaptation challenges.

\section{Related Work}\label{sec:background}

\textbf{DDPM:} Let $\mby_0 \sim q(\mby_0)$ be a data sample. The denoising diffusion probabilistic model (DDPM) \cite{ho2020denoising} involves the gradual addition of noise to $\mby_0$ in the forward process and then learning how to recover it in the reverse process. Specifically, in the forward process, $\mby_n$ at each time step $n \in \![\![1, N\!]\!]$ is derived via $q(\mby_n\mid \mby_{n-1})=\mathcal{N}(\mby_n; \sqrt{1-\beta_{n}} \mby_{n-1})$, where $(\beta_n)_n$ is a variance schedule and $N$ is the maximum number of time steps. Therefore we have:
\begin{align}
\label{eqn:diff_proc_forward}
\mby_0 \sim q (\mby_0), \quad &q\left(\mby_{1:N} \mid \mby_0\right) \triangleq \prod_{n=1}^{N}q\left(\mby_n\mid \mby_{n-1}\right).
\end{align}
Denoting $\alpha_n\triangleq 1-\beta_n, \bar{\alpha}_n\triangleq\prod_{s=1}^n \alpha_s$ and ${\mbep} \sim \mathcal{N}(\mbep;\mathbf{0}, \mbI)$, $\mby_n$ can be obtained in closed form for any given time $n$ as follows:
\begin{equation}
\label{eq:closed_form}
    \mby_n=\sqrt{\bar{\alpha}_n} \mby_0+\sqrt{1-\bar{\alpha}_n} {\mbep}.
\end{equation} 

In the reverse process, we want to reconstruct $\mby_0$. Since $q(\mby_{n-1}|\mby_{n})$ is intractable, a neural network model is trained to match the true denoising distribution. The equation of the reverse process is given by:
\begin{align}
\label{eqn:diff_proc_reverse}
p_{\theta}(\mby_{0:N})\triangleq p(\mby_N)\prod_{n=1}^{N}p_\theta\left(\mby_{n-1}\mid \mby_{n}\right), \quad 
\end{align}
where, assuming $N$ is large enough, $p_{\theta}(\mby_{n-1} | \mby_{n})$ is modeled as a Gaussian distribution with mean and variance determined by the neural network, and $\theta$ represents the parameters of the model, and $p(\mby_N) = \mathcal{N}(\mby_N; \mathbf{0}, \mbI)$.

In practice, the neural network, denoted by $\mbep_\theta(\mby_n, n)$, is parameterized to estimate the noise added to $\mby_0$ in Equation \ref{eq:closed_form}, and thus minimizes the following training objective, which bears resemblance to the concept of denoising score-matching:
\begin{equation}
    \mathbb{E}_{n, \mbep}\left[\left\|\mbep_\theta\left(\sqrt{\bar{\alpha}_n} \mby_0+\sqrt{1-\bar{\alpha}_n} \mbep, n\right)-\mbep\right\|_2^2\right]. \label{eq:diff_model}
\end{equation}

\textbf{WaveGrad:}  WaveGrad \cite{chen2020wavegrad} is based on DDPM and proposes to perform waveform generation with a conditional diffusion process $p_\theta(\mby_{0:n} \mid \tilde{\bold{X}})$ where $\tilde{\bold{X}}$ is a mel spectrogram.
Starting from a Gaussian white noise signal, it progressively enhances the signal through an iterative refinement process using a gradient-based sampler that takes into account $\tilde{\mbX}$ as a conditioning factor.
WaveGrad provides a flexible mechanism for balancing inference speed against sample quality by controlling the number of refinement steps. 
In contrast with DDPM~\cite{ho2020denoising}, WaveGrad \cite{chen2020wavegrad} reparameterized the model to condition on a continuous noise level $\bar \alpha$ instead of the discrete iteration index $n$. The loss function is then modified as follows:
\begin{equation}
    \mathbb{E}_{\bar{\alpha}, \mbep}\left[\left\|\mbep_\theta\left(\sqrt{\bar{\alpha}} \mby_0+\sqrt{1-\bar{\alpha}} \mbep, \tilde{\mbX}, \sqrt{\bar{\alpha}}\right)-\mbep\right\|_1\right].
\end{equation}
Considering $\mbz \sim \mathcal{N}(\mbz; \mathbf{0}, \mathbf{I})$ for $n > 1$, $\mbz = \mathbf{0}$ for $n=1$ and $\sigma_n = \frac{1 - \bar{\alpha}_{n - 1}}{1 - \bar{\alpha}_n}\beta_n$, Wavegrad applies the following iterative procedure to generate $\mby_{n-1}$:
\begin{equation}
\mby_{n-1}=\frac{\left(\mby_n-\frac{1-\alpha_n}{\sqrt{1-\bar{\alpha}_n}} \mbep_\theta\left(\mby_n, \tilde{\mbX}, \sqrt{\bar{\alpha}_n}\right)\right)}{\sqrt{\alpha_n}} + \sigma_n \mbz. \label{eq:wavegrad_gen}
\end{equation}

\textbf{SpecGrad:} Based on WaveGrad, SpecGrad \cite{koizumi2022specgrad} proposed to adjust the diffusion noise in a way that aligns its dynamic spectral characteristics with those of the conditioning mel spectrogram. The adaptation performed by SpecGrad through time-varying filtering results in notable enhancements in audio fidelity, particularly in the higher frequency regions. SpecGrad uses the implementation of a time-varying filter within the time-frequency domain.
We denote by $\mbT$ a matrix representation of the short-time Fourier transform (STFT) expressed from the time-domain to a flattened version of the time-frequency domain where all frames are concatenated, and by $\mbT^{\dagger}$ the similar matrix representation of the inverse STFT (iSTFT).
The time-varying filter can be expressed as: 
\begin{equation}
    \mbL=\mbT^{\dagger} \mbD \mbT,
\end{equation}
where $\mbD$ is a diagonal matrix determining the filter in the time-frequency domain, here obtained from the spectral envelope.
From $\mbL$, we can obtain the covariance matrix $\mbSi=\mbL\mbL^\Tr$ of the Gaussian noise 
$\mathcal{N}(\mathbf{0},\mbSi)$ in the diffusion process.
The loss function is then obtained as:
\begin{equation}
    \mathbb{E}_{\bar{\alpha}, \mbep}
    \left[
    \left\|\mbL^{-1}\left(\mbep_\theta\left(\sqrt{\bar{\alpha}} \mby_0+\sqrt{1-\bar{\alpha}} \mbep, \tilde{\mbX}, \sqrt{\bar{\alpha}}\right)-\mbep \right)\right\|_2^2\right]. \\
\end{equation}

\section{Proposed Griffin-Lim Diffusion Extension}
\label{sec:GLA-Grad}

\subsection{Griffin-Lim Algorithm}

The Griffin-Lim Algorithm (GLA) \cite{griffin1984signal} iteratively reconstructs a time-domain signal $s\in\RR^{L}$ of length $L$ from a given magnitude spectrogram $\hat{\mbS}\in \RR^{T\times F}$ by estimating a phase that is most consistent with that magnitude \cite{LeRoux2008SAPA09b}. $T$ and $F$ denote the time and frequency dimensions of the time-frequency (TF) representation.
For simplicity, we use the same notations $\mbT$ and $\mbT^{\dagger}$ for the STFT and iSTFT operators as for their matrix representations introduced earlier.
GLA relies on two projection operations in the complex time-frequency domain $\CC^{T\times F}$,
the first onto the subset $\mathcal{C}\subset \CC^{T\times F}$ of consistent spectrograms (the elements of  $\CC^{T\times F}$ that can be obtained as STFT of a signal in $\RR^{L}$):
\begin{equation}
P_{\mathcal{C}}(\mbC)=\mbT\mbT^{\dagger} \mbC,
\label{eq:Pc1}
\end{equation}
and the second onto the subset $\{\mbC \in \CC^{T\times F} \mid |\mbC|=\hat{\mbS} \}$ of spectrograms with magnitude equal to $\hat{\mbS}$: 
\begin{equation} 
P_{|\cdot|=\hat{\mbS}}(\mbC)=\hat{\mbS} \odot \frac{\mbC}{\left|\mbC\right|},
\label{eq:Pc2}
\end{equation} 
where $\odot, \frac{\cdot}{\cdot}$ are element-wise product and division respectively. 
The algorithm is initialized with $\mbC_0=\hat{\mbS}\odot e^{i\mbPh_0}$ where $\mbPh_0$ is typically random.
The $k$-th iteration can be summarized as: 
\begin{equation}
\mbC_k=P_{\mathcal{C}} \circ P_{|\cdot|=\hat{\mbS}}\left(\mbC_{k-1}\right).
\label{eq:GLA}
\end{equation}
Note that in practice, we use the Fast Griffin-Lim Algorithm (FGLA) \cite{perraudin2013fast}, a slightly modified version of GLA which accelerates the optimization and can be readily integrated in the modifications of the diffusion model based on GLA described below.

\subsection{GLA-Grad: Griffin-Lim corrected diffusion model}
We propose to introduce an inference-time improvement to a trained WaveGrad model. 
In current diffusion models such as WaveGrad, the iteration process (cf.\ Eq.~\eqref{eq:wavegrad_gen}) may lead to a signal $\mby_{n-1}$ that is out of distribution of the training data if not carefully considered, as introduced in \cite{graham2023denoising}. We believe that only conditioning on $\tilde{\mbX}$ may not be sufficient to solve the problem and a stronger constraint could be beneficial. 
 Our goal is to use GLA to fix the bias between the generated signal and our expected signal (see Fig.~\ref{fig:method}). 
\if 0 
We propose to introduce GLA into each step of the diffusion process, in which 
an estimate of the desired magnitude spectrogram  is used to guide the generation of the signal estimate $\mby_{n-1}$ at step $n-1$,
enabling the inference to have a better convergence.
 
Because the signal at step $n-1$ of the forward diffusion process is supposed to be obtained from a scaled version of the original $\mby_0$, the part of the magnitude $|\mbT \mby_{n-1}|$ that corresponds to the desired signal is a rescaled version of the desired magnitude as well: 

\begin{equation}
        \left|\mbT\mby_{n-1}\right| = \left|\sqrt{\bar{\alpha}_{n-1}}\mbT  \mby_0+ \sqrt{\left(1-\bar{\alpha}_{n-1}\right)} \mbT  \mbep \right|,
\label{equ-Tyn}
\end{equation}
with the signal component's magnitude corresponding to $\sqrt{\bar{\alpha}_{n-1}}|\mbT  \mby_0| $.

We do not have access to the desired magnitude $|\mbT \mby_0|$, but can consider the magnitude spectrogram estimate $\hat{\mbS}$ obtained from the desired mel spectrogram $\tilde{\mbX}$ 
through the pseudo-inverse $\mbM^\dagger$ of the mel transform as $\hat{\mbS} = \mbM^\dagger \tilde{\mbX}$. 

At diffusion time step $n-1$, we can then use GLA with the rescaled magnitude spectrogram $\hat{\mbS}_{n-1}$ as desired magnitude:
\begin{equation}
\hat{\mbS}_{n-1} = \sqrt{\bar{\alpha}_{n-1}}\hat{\mbS}.
\end{equation}

Our approach aims to minimize the disparity between the magnitude of the generated signal $|\mbT\mby_{n-1}|$ and the estimate  $\hat{\mbS}_{n-1}$ of the desired magnitude spectrogram at every iteration. 

We replace the original $P_{|\cdot|=\hat{\mbS}}$ projection given in Eq.~\eqref{eq:Pc2} by a new projection $P_{|\cdot|=\hat{\mbS}_{n-1}}$ onto the subset of spectrograms whose magnitude is $\hat{\mbS}_{n-1}$:

\begin{equation}
\begin{split}
    P_{|\cdot|=\hat{\mbS}_{n-1}}(\mbC)&=\hat{\mbS}_n \odot \frac{\mbC}{\left|\mbC\right|} =\sqrt{\bar{\alpha}_{n-1}}\mbM^\dagger\tilde{\mbX} \odot\frac{\mbC}{\left|\mbC\right|}. \\
\end{split}
\label{equ-QC2}
\end{equation}

Starting from the current diffusion estimate $\mby_{n-1}$ obtained using WaveGrad with Eq.~\eqref{eq:wavegrad_gen}, we can introduce a Griffin-Lim correction leading to a new estimate $\mby_{n-1}^{\text{GLA}}$ defined as:
\begin{equation}
\begin{split}
   
   \mby_{n-1}^{\text{GLA}} & =\mbT^{\dagger} (P_{\mathcal{C}} \circ  P_{|\cdot|=\hat{\mbS}_{n-1}})^K \left(\mbT\mby_{n-1} \right),
   \label{eq:y_gla}
\end{split}
\end{equation}
where $(P_{\mathcal{C}} \circ  P_{|\cdot|=\hat{\mbS}_{n-1}})^K$ indicates the application of $K$ iterations of GLA, starting from the spectrogram $\mbT\mby_{n-1}$ of $\mby_{n-1}$ as initial value.
\else We propose to introduce GLA into each step of the diffusion process, in which an estimate of the desired magnitude spectrogram  is used to guide the generation of the signal estimate $\mby_{n-1}$ at step $n-1$ by minimizing the disparity with its magnitude $|\mbT\mby_{n-1}|$, enabling the inference to have a better convergence.
We do not have access to the desired magnitude, but can consider the magnitude spectrogram estimate $\hat{\mbS}$ obtained from the desired mel spectrogram $\tilde{\mbX}$ 
through the pseudo-inverse $\mbM^\dagger$ of the mel transform as $\hat{\mbS} = \mbM^\dagger \tilde{\mbX}$. 
At diffusion time step $n-1$, we can then use GLA with the magnitude spectrogram $\hat{\mbS}$ as desired magnitude.

Starting from the current diffusion estimate $\mby_{n-1}$ obtained using WaveGrad with Eq.~\eqref{eq:wavegrad_gen}, we can introduce a Griffin-Lim correction leading to a new estimate $\mby_{n-1}^{\text{GLA}}$ defined as:
\begin{equation}
\begin{split}
   \mby_{n-1}^{\text{GLA}} & =\mbT^{\dagger} (P_{\mathcal{C}} \circ  P_{|\cdot|=\hat{\mbS}})^K \left(\mbT\mby_{n-1} \right),
   \label{eq:y_gla}
\end{split}
\end{equation}
where $(P_{\mathcal{C}} \circ  P_{|\cdot|=\hat{\mbS}})^K$ indicates the application of $K$ iterations of GLA, starting from the spectrogram $\mbT\mby_{n-1}$ of $\mby_{n-1}$ as initial value.
\fi

In practical implementation, our algorithm comprises two distinct phases, as depicted in Fig. \ref{fig:method}, showcasing the entire workflow. The initial phase employs the GLA-corrected approach outlined in this subsection at each sampling step of the diffusion process. More precisely, we obtain $\mby_{n-1}$ using the original WaveGrad update in Eq.~\eqref{eq:wavegrad_gen}, then correct it with GLA to obtain $\mby_{n-1}^{\text{GLA}}$ using Eq.~\eqref{eq:y_gla}. We posit that commencing the diffusion process with a projection method like GLA aids in maintaining signal consistency. Subsequently, the second phase involves a conventional diffusion process, exclusively considering the diffusion update. The proposed overall scheme is referred to as GLA-Grad.

\section{Evaluation}\label{sec:experiments}
\subsection{Datasets and Metrics}

For our experiments, we selected two widely used speech datasets, namely: 
\begin{itemize} 
    \item \textit{LJ Speech dataset} \cite{ljspeech17}, a public collection of 13,100 short audio clips featuring a single speaker reading passages from 7 non-fiction books. 
    The duration of the clips ranges from 1 to 10 seconds, resulting in a total length of approximately 24 hours. The sampling frequency is $22050$ Hz. We employ $150$ files for testing and allocate the remainder for training.
    \item \textit{Centre for Speech Technology Voice Cloning Toolkit (VCTK)} dataset \cite{Yamagishi2019CSTRVC}, a speech dataset spoken by 109 native English speakers representing various accents.
    Each speaker was prompted to read approximately 400 sentences. 
    We resample the VCTK dataset at the same sample rate as LJ Speech.
\end{itemize}
For all datasets, we use an STFT with $2048$ FFT points, a hop size of $300$, and a Hann window of length $1200$.
For the mel filterbank, $128$ filterbanks were considered with a lower frequency cutoff at $20$ Hz. The mel spectrograms were computed from the ground-truth audio.

The three metrics used for evaluation are: 
\begin{itemize} 
    \item \textit{Perceptual Evaluation of Speech Quality (PESQ)} \cite{pesq},
    a family of standards comprising a test methodology for automated assessment of speech quality as experienced by a user of a telephony system. 
    \item \textit{Short Term Objective Intelligibility measure (STOI)} \cite{stoi}, a metric to predict intelligibility of (quite) noisy speech.
    \item \textit{WARP-Q} \cite{warpQ}, an objective speech quality metric based on a subsequence dynamic time warping (SDTW) algorithm which yields a raw quality score of the similarity between the ground truth and the generated speech signal.
\end{itemize}

\subsection{Methods}
The four models considered for this evaluation include our GLA-Grad model and three baseline algorithms, namely WaveGrad, SpecGrad, and the original Griffin-Lim Algorithm (GLA) applied to the magnitude spectrogram obtained as pseudo-inverse of the conditioning mel spectrogram. 
We provide further implementation details on all four algorithms below.

\textbf{WaveGrad:} We train and evaluate a WaveGrad Base model \cite{chen2020wavegrad}. The noise schedule is set to the WaveGrad 6-step schedule, which is denoted as WG-6 and specified as 
$[\num{7e-6}, \num{1.4E-4}, \num{2.1E-3}, \num{2.8E-2}, \num{3.5E-1}, \num{7E-1}]$. 
This schedule is also employed and evaluated by SpecGrad \cite{koizumi2022specgrad}.
The rest of the parameters (batch size, sample rate) are all set as given in \cite{chen2020wavegrad} for the original WaveGrad model. We also report results with the much longer 50-step inference schedule WG-50  \cite{chen2020wavegrad}, to investigate whether a longer reverse diffusion process can lead to better results, at the expense of speed. This latter model is referred to as WaveGrad-50 in the results.

\textbf{SpecGrad:} For fair comparison, the same parameters as WaveGrad are used for training SpecGrad, and we keep the same noise-shaping parameters as in the original SpecGrad algorithm. 
SpecGrad actually introduced conditioning information into the sampling process since the noise-shaping could be regarded as a reconstruction. 
Its goal is thus somehow aligned with our approach for minimizing the conditioning error. The same WG-6 inference schedule is used, as in \cite{koizumi2022specgrad}.

\textbf{Griffin-Lim Algorithm:} We employ the Fast Griffin-Lim Algorithm \cite{perraudin2013fast} for phase reconstruction, whose implementation is available in torchaudio. A total of 1000 iterations is used to estimate the phase and get the reconstructed signal. 

\textbf{GLA-Grad}: GLA-Grad itself does not require training, and is built on top of a WaveGrad model trained as described above. The same WG-6 noise schedule 
is used in this experiment for comparing with WaveGrad and SpecGrad. We find that the alternative projection cannot keep correcting the signal in the entire reverse process. On the contrary, the projection may be harmful when the quality of the generated wave is high enough. Thus, we only apply the correction at the first $3$ reverse steps, then using the unmodified diffusion update for the remaining steps.
A total of $K=32$ alternative projections are used in Eq.~\ref{eq:y_gla}  for the correction during each reverse step in the first phase.  
We use the pseudo-inverse of the mel matrix to convert mel spectrogram to magnitude STFT.

Audio examples as well as the full code of our model released as open-source software are available on our demo page\footnote{\url{https://GLA-Grad.github.io/}}.
 
\subsection{Experiments and discussion}

We propose several experiments 
with the objective to assess the performance and effectiveness of the proposed approach in addressing generation efficiency, in particular for speech generation of unseen speakers at training time. 

\noindent We conducted experiments under three setups, training WaveGrad and SpecGrad and estimating all methods in each scenario:
\begin{itemize}
    \item \textit{Closed single speaker:} We train and evaluate on the LJ Speech Dataset, using the isolated validation set for test. 
    \item \textit{Generalization from a single speaker:} The models trained on LJ Speech as above are evaluated on the VCTK multi-speaker dataset for testing the generalization capability of the model from a single speaker's data. We select 19 speakers from VCTK and use 150 utterances for evaluation.
    \item \textit{Adaptation to new speakers:} We train and evaluate the models on VCTK, for further evaluating their domain adaption ability when given a larger variety of training data. The speech recordings of 19 Speakers (7400 utterances in total) are used as the training set, while we pick 150 utterances from the remaining 90 speakers for evaluation.
\end{itemize}

\begin{table}[t]
\centering
    \sisetup{
    detect-weight, 
    mode=text, 
    tight-spacing=true,
    round-mode=places,
    round-precision=3,
    table-format=1.3,
    table-number-alignment=center
    }    
        \caption{Results when training and evaluating on LJ Speech}
    \resizebox{0.97\columnwidth}{!}{
    \begin{tabular}{l
    S[round-precision=2,table-format=1.2]@{\,\( \pm \)\,}S[round-precision=2,table-format=1.2]
    *{2}{S@{\,\( \pm \)\,}S}}
    \toprule
    {\bfseries Model} & \multicolumn{2}{c}{\bfseries PESQ ($\uparrow$)} & \multicolumn{2}{c}{\bfseries STOI ($\uparrow$)} & \multicolumn{2}{c}{\bfseries WARP-Q ($\downarrow$)}\\ \midrule
        GLA-Grad & 3.460 & 0.112 & 0.963 & 0.005  &  1.677&0.076 \\ 
        WaveGrad  & 3.592&0.128 & 0.970&0.004 & 1.654 & 0.075\\
        WaveGrad-50  & \bfseries 3.72&0.110 & \bfseries 0.978&0.004 & \bfseries 1.363&0.054 \\
        SpecGrad  & 3.618&0.142 & 0.963&0.005 & 1.408 & 0.054\\
        Griffin-lim  & 1.023&0.004 & 0.565&0.042 & 3.234 & 0.118\\
    \bottomrule
    \end{tabular}
    }
    \label{tab:Experiment1}
\end{table}

\begin{table}[t]
   \centering
       \sisetup{
    detect-weight, 
    mode=text, 
    tight-spacing=true,
    round-mode=places,
    round-precision=3,
    table-format=1.3,
    table-number-alignment=center
    }    
     \caption{Results when training on LJ Speech and evaluating on 19 speakers of VCTK}
    \resizebox{0.97\columnwidth}{!}{
    \begin{tabular}{l
    S[round-precision=2,table-format=1.2]@{\,\( \pm \)\,}S[round-precision=2,table-format=1.2]
    *{2}{S@{\,\( \pm \)\,}S}}
    \toprule
    {\bfseries Model} & \multicolumn{2}{c}{\bfseries PESQ ($\uparrow$)} & \multicolumn{2}{c}{\bfseries STOI ($\uparrow$)} & \multicolumn{2}{c}{\bfseries WARP-Q ($\downarrow$)}\\ \midrule
         GLA-Grad & \bfseries 2.734 & 0.283 &\bfseries 0.944&0.017  &  1.722&0.132 \\ 
         WaveGrad  & 2.076&0.310 & 0.873&0.035 & 1.913&0.128 \\
         WaveGrad-50  & 1.997&0.293 & 0.670&0.111 & 2.122&0.411 \\
         SpecGrad  & 2.481&0.380  & 0.812&0.066 & \bfseries 1.593&0.103\\
         Griffin-lim & 1.036&0.013 & 0.522&0.098 & 3.411&0.164 \\
    \bottomrule
    \end{tabular}
    }
    \label{tab:Experiment2}
\end{table}

WaveGrad and SpecGrad slightly outperform GLA-Grad on LJ Speech (see Table \ref{tab:Experiment1}), which confirms their performance on speakers known at training time. In our experiments, we observed that the performance of our model varies with the number of GLA steps in inference, especially for the  
steps near $\mby_0$. An optimisation of this amount of GLA steps should help the model be more stable. Note though that GLA-Grad has the lowest standard deviation on PESQ, indicating a superior stability in output quality. 

When considering generalization from a single speaker, GLA-Grad outperforms both WaveGrad and SpecGrad (see Table \ref{tab:Experiment2}). SpecGrad outperforms other models on WARP-Q metric, while GLA-Grad has far higher score on PESQ and STOI. As in the closed speaker setup, 
GLA-Grad has the lowest standard deviation on both PESQ and STOI. It can be observed that the performance of WaveGrad and SpecGrad drops dramatically due to the external evaluation data set which includes more speakers, unseen at training, whereas GLA-Grad sees a much smaller decrease in performance.
The results for adaptation to new speakers on VCTK, presented in Table \ref{tab:Experiment3}, show that GLA-Grad also outperforms other models for a larger unseen speaker set, and that its generalization capability is not limited to the extreme scenario where only data from a single speaker is used for training. 

Looking at the results of the more computationally intensive WaveGrad-50 across the various settings, we see that it does improve performance in the closed single-speaker setting, but actually resulted in a significantly decreased performance in the open speaker settings, the longer inference process harming the generalization ability of the WaveGrad model.

In summary, it can be noticed that, while WaveGrad and SpecGrad perform best in the closed single-speaker setting, 
GLA-Grad clearly outperforms all baselines in cross-domain situations where there exists important speaker differences between training and evaluation sets. 
This indicates that our Griffin-Lim corrected WaveGrad model exhibits stronger generalization performance to speakers who are unseen at training, without needing fine-tuning or modification to the training. 

\begin{table}[t]
\centering
    \sisetup{
    detect-weight, 
    mode=text, 
    tight-spacing=true,
    round-mode=places,
    round-precision=3,
    table-format=1.3,
    table-number-alignment=center
    }    
     \caption{Results when training on 19 speakers of VCTK and evaluating on 90 other speakers of VCTK}
     \resizebox{0.97\columnwidth}{!}{
    \begin{tabular} {l
    S[round-precision=2,table-format=1.2]@{\,\( \pm \)\,}S[round-precision=2,table-format=1.2]
    *{2}{S@{\,\( \pm \)\,}S}}
    \toprule
    {\bfseries Model} & \multicolumn{2}{c}{\bfseries PESQ ($\uparrow$)} & \multicolumn{2}{c}{\bfseries STOI ($\uparrow$)} & \multicolumn{2}{c}{\bfseries WARP-Q ($\downarrow$)}\\ \midrule
         GLA-Grad & \bfseries 2.883&0.44 & \bfseries 0.856&0.081 & 1.520&1.102 \\ 
         WaveGrad  & 2.093&0.476 & 0.803&0.076 & 1.801&0.133 \\
         WaveGrad-50  & 1.994&0.376 & 0.706&0.093 & 2.024&0.157 \\
         SpecGrad  & 2.563&0.354 & 0.814&0.080 & \bfseries 1.492& 0.127 \\
         Griffin-lim  & 1.036&0.019 &0.542&0.112 & 3.410&0.169\\
    \bottomrule
    \end{tabular}
    }
    \label{tab:Experiment3}
\end{table}

\subsection{Complexity}

Table \ref{tab:ComplexityStudy} shows the inference speed compared to real-time evaluated with 150 utterances in our 3 setups on an NVIDIA A100 GPU. In terms of complexity, the proposed model is comparable in speed with SpecGrad on LJ Speech and somewhat slower on VCTK, while the difference with the baseline WaveGrad model is higher. The computation of the pseudo-inverse of the mel spectrogram accounts for a significant amount of the speed difference between the methods, while the extra Griffin-Lim steps performed at a given diffusion step are not that computationally intensive, and are in fact faster than the diffusion model sampling iteration itself. Adding these Griffin-Lim iterations in GLA-Grad does reduce speed, but it is both an effective and efficient way to improve performance, as adding more diffusion steps has a higher cost in terms of speed, while potentially leading to worse performance as shown in the results of WaveGrad-50.
The proposed GLA-Grad thus provides a good trade-off between quality and inference speed.

\begin{table}[t]
\centering
    \sisetup{
    detect-weight, 
    mode=text, 
    tight-spacing=true,
    round-mode=places,
    round-precision=1,
    table-format=2.1,
    table-number-alignment=center
    }
    \caption{Inference speed compared to real-time ($\uparrow$)}
    \setlength{\tabcolsep}{5pt}
    \resizebox{0.97\columnwidth}{!}{
    \begin{tabular}{l  *{3}{S}}
    \toprule
         {\bfseries Model} & 
         {\bfseries LJ$\to$LJ} &
         {\bfseries LJ$\to$VCTK} &
         {\bfseries VCTK$\to$VCTK} \\ 
    \midrule
         GLA-Grad & 39.04 & 25.1 & 24.06\\ 
         WaveGrad  & \bfseries 54.28 & \bfseries 55.73 & \bfseries 56.05 \\
         WaveGrad-50 & 7.28 & 7.45 & 7.46\\
         SpecGrad & 40.51 & 42.71 & 43.29 \\
         Griffin-lim & 26.35 & 11.62 & 10.81 \\
    \bottomrule
    \end{tabular}
    }
    \label{tab:ComplexityStudy}
\end{table}

\section{Conclusion \& Future Works}\label{sec:conclusion}
In this paper, we proposed GLA-Grad, a novel scheme for diffusion-based generation of speech from mel spectrogram. Our experiments show in particular that our model clearly outperforms several baselines for unseen speakers at training. These results are obtained by refining the spectrogram phase for the first steps of the diffusion process using a Griffin-Lim algorithm. 
This shows that it is beneficial to incorporate a phase retrieval module within the diffusion process, which opens new perspectives for diffusion-based speech generation. Future work will be dedicated to the extension of our model with recent and more accurate phase retrieval methods which should further help for speech generation of unseen speakers. 

\balance
\bibliographystyle{IEEEtran}
\bibliography{refs}

\end{document}